\documentclass[twoside,twocolumn,9pt]{article}
\usepackage{extsizes}
\usepackage[super,sort&compress,comma]{natbib} 
\usepackage[version=3]{mhchem}
\usepackage[left=1.5cm, right=1.5cm, top=1.785cm, bottom=2.0cm]{geometry}
\usepackage{balance}
\usepackage{mathptmx}
\usepackage{sectsty}
\usepackage{graphicx} 
\usepackage{lastpage}
\usepackage[format=plain,justification=justified,singlelinecheck=false,font={stretch=1.125,small,sf},labelfont=bf,labelsep=space]{caption}
\usepackage{float}
\usepackage{fancyhdr}
\usepackage{fnpos}
\usepackage[english]{babel}
\addto{\captionsenglish}{%
  
}
\usepackage{array}
\usepackage{droidsans}
\usepackage{charter}
\usepackage[T1]{fontenc}
\usepackage[usenames,dvipsnames]{xcolor}
\usepackage{setspace}
\usepackage[compact]{titlesec}
\usepackage{hyperref}
\usepackage{bm}

\usepackage{epstopdf}

\definecolor{cream}{RGB}{222,217,201}

\begin{document}

\pagestyle{fancy}
\thispagestyle{plain}
\fancypagestyle{plain}{
\renewcommand{\headrulewidth}{0pt}
}

\makeFNbottom
\makeatletter
\renewcommand\LARGE{\@setfontsize\LARGE{15pt}{17}}
\renewcommand\Large{\@setfontsize\Large{12pt}{14}}
\renewcommand\large{\@setfontsize\large{10pt}{12}}
\renewcommand\footnotesize{\@setfontsize\footnotesize{7pt}{10}}
\makeatother

\renewcommand{\thefootnote}{\fnsymbol{footnote}}
\renewcommand\footnoterule{\vspace*{1pt}%
\color{cream}\hrule width 3.5in height 0.4pt \color{black}\vspace*{5pt}} 
\setcounter{secnumdepth}{5}

\makeatletter 
\renewcommand\@biblabel[1]{#1}            
\renewcommand\@makefntext[1]%
{\noindent\makebox[0pt][r]{\@thefnmark\,}#1}
\makeatother 
\renewcommand{\figurename}{\small{Fig.}~}
\sectionfont{\sffamily\Large}
\subsectionfont{\normalsize}
\subsubsectionfont{\bf}
\setstretch{1.125} 
\setlength{\skip\footins}{0.8cm}
\setlength{\footnotesep}{0.25cm}
\setlength{\jot}{10pt}
\titlespacing*{\section}{0pt}{4pt}{4pt}
\titlespacing*{\subsection}{0pt}{15pt}{1pt}

\fancyfoot{}
\fancyfoot[RO]{\footnotesize{\sffamily{1-\pageref{LastPage} ~\textbar  \hspace{2pt}\thepage}}}
\fancyfoot[LE]{\footnotesize{\sffamily{\thepage~\textbar\hspace{3.45cm}}}}
\fancyhead{}
\renewcommand{\headrulewidth}{0pt} 
\renewcommand{\footrulewidth}{0pt}
\setlength{\arrayrulewidth}{1pt}
\setlength{\columnsep}{6.5mm}
\setlength\bibsep{1pt}

\makeatletter 
\newlength{\figrulesep} 
\setlength{\figrulesep}{0.5\textfloatsep} 

\newcommand{\topfigrule}{\vspace*{-1pt}%
\noindent{\color{cream}\rule[-\figrulesep]{\columnwidth}{1.5pt}} }

\newcommand{\botfigrule}{\vspace*{-2pt}%
\noindent{\color{cream}\rule[\figrulesep]{\columnwidth}{1.5pt}} }

\newcommand{\dblfigrule}{\vspace*{-1pt}%
\noindent{\color{cream}\rule[-\figrulesep]{\textwidth}{1.5pt}} }

\makeatother

\twocolumn[
  \begin{@twocolumnfalse}

\noindent\LARGE{\textbf{Acoustic Cavitation Rheometry}} \\
 \vspace{0.3cm} \\

 \noindent\large{Lauren Mancia,$^{\ast}$\textit{$^{a}$}  Jin Yang,\textit{$^{b}$} Jean-Sebastien Spratt,\textit{$^{c}$} Jonathan R. Sukovich,\textit{$^{d}$} Zhen Xu,\textit{$^{d}$} Tim Colonius,\textit{$^{c}$} Christian Franck,\textit{$^{b}$} and Eric Johnsen \textit{$^{a}$}} \\
 
\vspace{0.3cm}  

 \noindent\normalsize{Characterization of soft materials is challenging due to their high compliance and the strain-rate dependence of their mechanical properties.  The inertial microcavitation-based high strain-rate rheometry (IMR) method [Estrada \textit{et al., J. Mech. Phys. Solids}, 2018, \textbf{112}, 291-317] combines laser-induced cavitation measurements with a model for the bubble dynamics to measure local properties of polyacrylamide hydrogel under high strain-rates from $10^3$ to $10^8$ s$^{-1}$. While promising, laser-induced cavitation involves plasma formation and optical breakdown during nucleation, a process that could alter local material properties before measurements are obtained. In the present study, we extend the IMR method to another means to generate cavitation, namely high-amplitude focused ultrasound, and apply the resulting acoustic-cavitation-based IMR to characterize the mechanical properties of agarose hydrogels. Material properties including viscosity, elastic constants, and a stress-free bubble radius are inferred from bubble radius histories in $0.3\%$ and $1\%$ agarose gels.  An  ensemble-based data assimilation is used to further help interpret the obtained estimates. The resulting parameter distributions are consistent with available measurements of agarose gel properties and with expected trends related to gel concentration and high strain-rate loading. Our findings demonstrate the utility of applying IMR and data assimilation methods with single-bubble acoustic cavitation data for measurement of viscoelastic properties.} \\


 \end{@twocolumnfalse} \vspace{0.6cm}

  ]


\renewcommand*\rmdefault{bch}\normalfont\upshape
\rmfamily
\section*{}
\vspace{-1cm}


\footnotetext{\textit{$^{a}$~Department of Mechanical Engineering, University of Michigan, Ann Arbor, MI, USA. E-mail: lamancha@umich.edu}}
\footnotetext{\textit{$^{b}$~Department of Mechanical Engineering, University of Wisconsin-Madison, WI, USA. }}
\footnotetext{\textit{$^{c}$~Division of Engineering and Applied Science, California Institute of Technology, Pasadena, CA, USA. }}
\footnotetext{\textit{$^{d}$~Department of Biomedical Engineering, University of Michigan, Ann Arbor, MI, USA. }}





\section{Introduction}
Characterization of soft materials such as polymers, hydrogels, biological tissues, and tissue phantoms is important to a variety of engineering and biomedical applications \cite{chaudhuri2016hydrogels,lee2012alginate,storrie2006sustained,solomon2007modeling}. Soft materials are challenging to characterize, particularly at high rates, given their inhomogeneity, high compliance \cite{arora1999compliance}, and the dependence of their mechanical properties on strain rate\cite{brujan2006stress}. Traditional methods measure bulk material properties under quasi-static loading conditions. At high strain-rates, dynamic loading tests such as the Taylor impact test \cite{taylor1948use,allen1997optimizing} and split Hopkinson pressure bar \cite{chen2010split} are typically used. The Taylor impact test emulates ballistic loading but is inherently destructive and poorly suited to soft, highly compliant materials. The split Hopkinson pressure bar is more versatile and has been used to characterize biological materials including bone \cite{kulin2011effects} and muscle \cite{van2006high}. However, its use is generally limited to loading rates of $10^4$ s$^{-1}$ and higher applied stresses. There are also technical challenges associated with specimen preparation \cite{chen2010split,hu2012indentation}. 

Using relatively simple and inexpensive setups, cavitation has enabled rheometry techniques capable of probing the local material properties of complex, soft specimens \cite{barney2020cavitation,zimberlin2007cavitation,estrada2018high}. The first of these methods, the cavitation rheology technique (CRT), involves creating a cavity in soft material via needle injection of air and measurement of the critical pressure corresponding to mechanical instability. The critical pressure is directly related to the material's elastic modulus. CRT has been successfully applied to the measurement of eye vitreous \cite{zimberlin2010cavitation}, eye lens \cite{cui2011cavitation}, skin \cite{chin2013cavitation}, and polymer \cite{bentz2016solvent} properties. This method is minimally invasive, cost-effective, efficient, and applicable at microscopic length scales; however, injection must be slow enough for a quasi-static assumption to hold \cite{zimberlin2007cavitation,estrada2018high}. The Volume Controlled Cavity Expansion (VCCE) method is a needle-based cavitation rheometry technique that permits inference of rate-dependent material properties without knowledge of the maximum recorded cavity pressure \cite{raayai2019volume}; however, \citet{chockalingam2020probing} note that VCCE is still limited to strain rates of $\sim1$ s$^{-1}$. Thus, both CRT and VCCE have limited ability to characterize soft materials at the high strain rates ($>10^3$ s$^{-1}$) most relevant to blast injury diagnostics and mitigation \cite{nyein2010silico,ramasamy2011blast}, focused ultrasound ablation \cite{mancia2019modeling}, and laser surgery \cite{brujan2006stress,vogel2008femtosecond}. Recently, novel cavitation-based rheometry techniques have been developed to characterize soft materials subjected to these extreme conditions \cite{estrada2018high,yang2020extracting}. Unlike traditional methods for high strain-rate material characterization \cite{chen2010split}, techniques such as Small-scale Ballistic Cavitation (SBC) \cite{milner2019device} and Inertial Microcavitation-based high strain-rate Rheometry (IMR) \cite{estrada2018high} share many of the advantages of CRT. The IMR method in particular has been used to measure nonlinear viscoelastic properties of polyacrylamide \cite{estrada2018high,yang2020extracting}.
 
 IMR uses high-speed videography to track the radius vs.\ time behavior of a bubble produced via inertial cavitation, then compares recorded radius measurements with numerical simulations that permit inference of viscoelastic material parameters for a given constitutive model. The method effectively characterizes the shear modulus and dynamic viscosity of polyacrylamide gels of varying stiffness \cite{estrada2018high}. IMR was first demonstrated with laser cavitation data, but the authors note that any input of energy capable of inducing inertial cavitation in the medium can be used to obtain radius vs.\ time measurements. Laser cavitation is initiated when the rapid concentration of high temperatures and pressures generated during laser plasma formation triggers explosive expansion \cite{vogel2008femtosecond}. In contrast, ultrasound generates cavitation when a sufficient pressure rarefaction causes some pre-existing defect or nucleus in a material to grow explosively into a larger cavity. Ultrasound-induced cavitation is not complicated by plasma formation, making it more relevant to focused ultrasound applications and potentially more analogous to tissue deformation in blast and ballistic injuries. However, given the technical difficulty of generating single bubbles with high-amplitude ultrasound, only recently have refined experimental techniques permitted a rigorous comparison of single-bubble dynamics generated via laser vs.\ ultrasound \cite{wilson2019comparative}. Acoustic cavitation data from \citet{wilson2019comparative} were subsequently used to measure the cavitation nucleus size distribution in water \cite{mancia2020measurements} and to validate existing models for single-bubble dynamics \cite{mancia2020single}.

The present study introduces acoustic cavitation rheometry as an extension of the IMR method for use with acoustic rather than laser-induced cavitation data. This technique is then used to characterize $0.3\%$ and $1\%$ agarose gel specimens first studied by \citet{wilson2019comparative}.  Bubble stress-free radius size and agarose properties, including viscosity and elastic constants, are inferred using a combination of ultrasound-induced bubble radius vs.\ time measurements and a numerical model for single bubble dynamics. The gel is modeled as a viscoelastic Kelvin-Voigt material with either a Neo-Hookean or strain-stiffening hyperelastic spring and a linear dashpot. Our parameter distributions are subsequently compared to available quasi-static measurements of agarose material properties. We then discuss additional sources of uncertainty  and provide a comparison of IMR applied with acoustic vs. laser-induced inertial cavitation data. We conclude with an analysis of the $0.3\%$ gel data using a recently proposed data assimilation modification of IMR \cite{Spratt_2020}, demonstrating the potential of acoustic cavitation rheometry to accommodate additional sources of modeling and experimental uncertainty.

\section{Methods}

\subsection{Experiments}
The experimental methods for generating single bubbles via high-amplitude ultrasound were described previously \cite{wilson2019comparative}. In this work, we analyze the $19$ data sets in $0.3\%$ agarose and the $20$ data sets in $1\%$ agarose from that study. To summarize, the gel specimens were prepared according to the procedure described by \citet{vlaisavljevich2015effects} with the modification that gels solidified at $17.8$ $^{\circ}$C rather than $4$ $^{\circ}$C. This slight difference in preparation had a negligible effect on the Young's moduli of each gel measured under quasi-static conditions \cite{vlaisavljevich2015effects}. Given that agarose can be considered incompressible at high strain rates \cite{normand2000new} with Poisson's ratio, $\nu \approx 0.5$, the shear modulus of each gel is taken to be 1/3 of its measured Young's modulus. Thus, the approximate quasi-static shear moduli of $0.3\%$ and $1\%$ agarose gels in this study are $0.38$ $\pm$ $0.16$ kPa and $7.2$ $\pm$ $0.33$ kPa, respectively. 

Experiments were performed in a open-topped, spherical acoustic array that was $10$\,cm in diameter and populated with $16$ focused transducer elements with a center frequency of $1$ MHz. A $5.8$\,cm-diameter opening at the top of the transducer permitted insertion of the gel specimens. Bubbles were nucleated using a $1.5$-cycle acoustic pulse containing a single rarefactional pressure half-cycle \cite{mancia2020measurements} with an amplitude of $-24$ MPa. For each experiment, bubbles were nucleated at least $5$ mm away from previous cavitation sites, and specimens had specific acoustic impedance close to that of water (within $5$\%) to ensure samples could be regarded as infinite relative to the bubbles. Bubbles were imaged through a single cycle of growth and collapse using a camera with a fixed frame rate of $400$ kHz. The multi-flash-per-camera-exposure technique \cite{sukovich2020cost} generated images of nested, concentric bubbles which were differentiated using brightness thresholding and edge detection. Bubble radii were measured at individual flash points by applying a circle fit to their detected boundaries. For all experiments, the magnitude of the spatial resolution uncertainty was less than 4.3 $\mu$m and temporal resolution uncertainty was less than 1.25 $\mu$s.

\begin{figure}[!t]
\centering
  \includegraphics[width=0.45\textwidth]{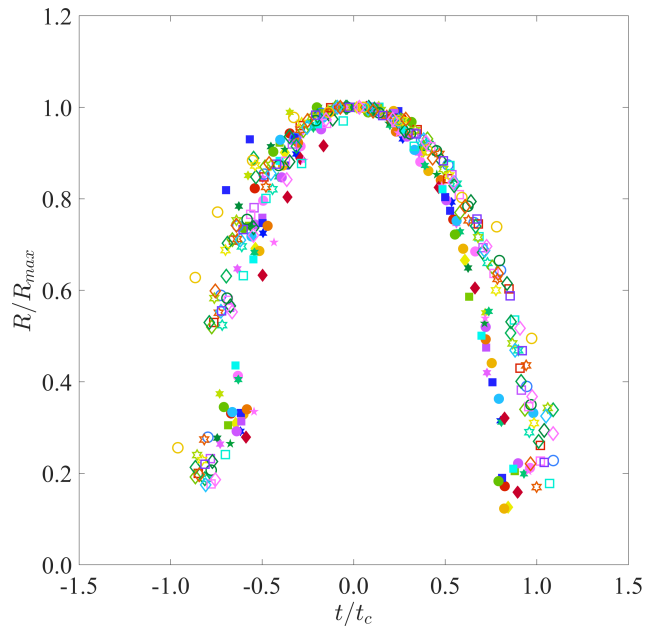}
  \caption{Scaled bubble radius vs.\ time data adapted from Wilson \textit{et al}. \cite{wilson2019comparative} for $19$ experiments in $0.3\%$ (open markers) agarose gel and $20$ experiments in $1\%$ (filled markers) agarose gel.}
  \label{fig:ScData}
\end{figure}

The scaled radius vs.\ time data, $(t_i,R_i)$ for all experiments is shown in Figure \ref{fig:ScData}, where the open markers correspond to the $0.3\%$ gel data and the filled markers correspond to the $1\%$ gel data. As in previous studies \cite{mancia2020measurements,wilson2019comparative,estrada2018high}, the scaling is by the maximum bubble radius, $R_{max}$ and the collapse time, $t_c$. As noted in a previous study of cavitation nuclei sizes in water \cite{mancia2020measurements}, this data collapse with appropriate scaling suggests that all experiments are governed by the same physics. In the present case, however, there are significant uncertainties in  the local material response and viscoelastic parameters of the gel specimens, as well as in the initial conditions. 

\subsection{Theoretical Model and Numerical Methods} \label{sec:model}

Numerical simulations are based on a theoretical model for  cavitation in a viscoelastic medium that has been used in multiple prior studies of ultrasound-induced cavitation \cite{mancia2019modeling,wilson2019comparative,bader2018influence,mancia2017predicting,vlaisavljevich2016effects,vlaisavljevich2016visualizing,vlaisavljevich2015effects} and is thus  described only briefly here. A spherical homobaric microbubble is subjected to a tensile half-cycle experimental waveform, $p_f(t)$, with an amplitude of $-24$ MPa. The Keller-Miksis equation \cite{keller1980bubble} is used to model spherical bubble dynamics in a homogeneous, weakly compressible medium:

\begin{align}
\label{KM}
\begin{split}
&\left(1 -\frac{\dot{R}}{c_{\infty}}\right)R\ddot{R}+\frac{3}{2}
\left(1 - \frac{\dot{R}}{3c_{\infty}}\right)\dot{R}^2 = \\
&\frac{1}{\rho_{\infty}}\left(1 +\frac{\dot{R}}{c_{\infty}}
+\frac{R}{c_{\infty}}\frac{d}{dt}\right)\Biggl[p_b-
\Biggl(p_{\infty}+ p_{f}\Biggl(t + \frac{R}{c_{\infty}}\Biggr)\Biggr)-\frac{2S}{R} + J_{SS} \Biggr],
\end{split}
\end{align}

\noindent
where the sound speed, $c_{\infty}$, density, $\rho_{\infty}$, far-field pressure, $p_{\infty}$, surface tension, $S$, and viscosity, $\mu$, are fixed at the values given by \citet{wilson2019comparative}  The time-dependent pressure inside the bubble, $p_b(t)$, is coupled to the energy equation solved inside the bubble  \cite{prosperetti1991thermal,prosperetti1988nonlinear,kamath1993theoretical}.  Gel surrounding the bubble remains at a constant ambient temperature of $25$ $^{\circ}$C, and the interface between the bubble and surrounding gel is assumed to be impervious to gas diffusion. This model has been used in previous studies  \cite{prosperetti1991thermal,prosperetti1988nonlinear,kamath1993theoretical,warnez2015numerical,mancia2020single}  of bubble growth and collapse observed in  acoustically nucleated cavitation experiments \cite{wilson2019comparative,warnez2015numerical,mancia2017predicting,barajas2017effects,mancia2020single}. 

The selection of an appropriate viscoelastic constitutive model for the gel specimens is nontrivial but can be elucidated with rigorous application of the IMR approach \cite{estrada2018high}. For this study, we model both gel specimens using Kelvin-Voigt-type models with either a Neo-Hookean \cite{gaudron2015bubble} or a higher-order strain-stiffening \cite{yang2020extracting} elastic term. The finite-deformation Neo-Hookean model was first applied in the context of inertial cavitation by \citet{gaudron2015bubble}. This model is favored for high-amplitude ultrasound simulations given the typically large bubble growth observed in these cases \cite{mancia2019modeling,mancia2017predicting,vlaisavljevich2016visualizing}. It was also used to model laser-induced inertial cavitation in polyacrylamide \cite{estrada2018high,yang2020extracting}. Additionally, higher-order strain-stiffening models \cite{movahed2016cavitation}, which are observed to be significant in the dynamic response of soft materials \cite{raayai2019intimate,raayai2019capturing,chen2011strain},  have recently been applied to model laser-induced inertial cavitation in polyacrylamide \cite{yang2020extracting}. Here we apply an adaptation of the Fung model \cite{fung2013biomechanics}, which is approximated by using the first two terms of the Taylor series expansion of the Fung model. For the remainder of the text we  refer to it as the Quadratic Law Kelvin-Voigt (QLKV) model \cite{yang2020extracting}. The integral of the deviatoric contribution of the stresses in the surrounding medium is given by:
\begin{equation}
\begin{aligned}
\label{eq:SS}
J_{SS}&= -\frac{4\mu \dot{R}}{R} + \frac{G(3\alpha - 1)}{2}\left[ 5 - 4
  \left(\frac{R_0}{R}\right) - \left(\frac{R_0}{R}
  \right)^4\right]\\
&\qquad + 2G\alpha\left[ \frac{27}{40} + \frac{1}{8}\left(\frac{R_0}{R}\right)^8 +\frac{1}{5}\left(\frac{R_0}{R} \right)^5+ \left(\frac{R_0}{R}\right)^2 -\frac{2R}{R_0} \right] ,
\end{aligned}
\end{equation}

\noindent
where $R$ is the time-dependent bubble radius and $R_0$ is the stress-free radius corresponding to a reference configuration; departures from this radius give rise to restoring elastic stresses. Constant viscoelastic properties of the gel specimens include gel viscosity, $\mu$, shear modulus, $G$, and a stiffening parameter, $\alpha$. Note that when $\alpha = 0$ the strain stiffening QLKV model reduces to the Neo-Hookean model. 

The above stress integral thus contains four parametric uncertainties ($G$, $\alpha$, $\mu$, and $R_0$), reducing to three for the Neo-Hookean case ($\alpha=0$). The stress-free radius, $R_0$, is taken to be the initial bubble radius in all simulations. Physical parameters for water and air are assumed constant for all simulations and are the same as those given in prior work\cite{wilson2019comparative,mancia2020measurements} with the exception of the material properties to be inferred. We adopt a previously presented non-dimensionalization for the resulting system of ODEs and PDE \cite{mancia2019modeling}. Time marching is achieved using a variable-step, variable-order solver based on numerical differentiation formulas (MATLAB \textit{ode15s}) \cite{shampine1997matlab,shampine1999solving}. Spatial derivatives in the energy equation are computed using second-order central differences \cite{estrada2018high,barajas2017effects}. 

\subsection{Inference of Material Properties}
\label{sec:inference}

\begin{figure}[h]
\centering
  \includegraphics[width = 0.45\textwidth]{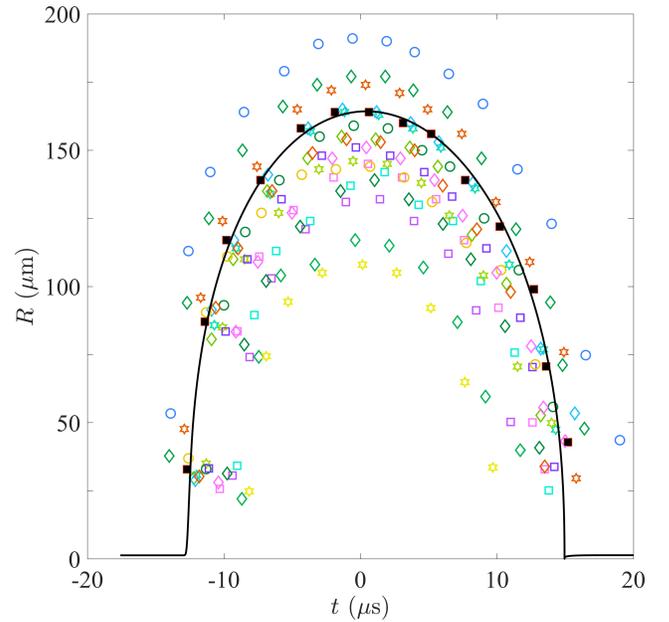}
  \caption{Dimensional radius vs.\ time data for $19$ experiments in $0.3\%$ agarose gel. The best fit simulation obtained with the Neo-Hookean model for a representative experimental data set (filled squares) is shown in black.}
  \label{fig:Dim03}
\end{figure}

\subsubsection{IMR Aproach}

Material properties for the $0.3 \%$ and $1 \%$ agarose gels are inferred using Neo-Hookean and QLKV viscoelastic models. The coupling of each experimental data set to uncertain material parameters is achieved with numerical simulations of single-bubble dynamics. For each experimental radius vs.\ time data set, we perform a series of simulations iterating over a maximum of four uncertain material parameters ($G$, $\alpha$, $\mu$, $R_0$) using one of three assumed models: (i) Neo-Hookean ($\alpha = 0$, 3 flexible parameters), (ii) quasi-static QLKV ($G$ fixed at measured quasi-static values for each gel concentration, 3 flexible parameters), and (iii) general QLKV (4 flexible parameters). Simulation results are time-shifted within $ \pm 0.8$ $\mu$s as needed for temporal alignment with a given experimental data set.  Next, the normalized root mean squared error (NRMSE) is calculated between the radius and time data points from a given experimental data set, $y_{exp}$, and their nearest neighbor points on a simulation trace, $y_{sim}$:

\begin{equation}
\begin{aligned}
\label{eq:NRMSE}
\mathrm{NRMSE} = 1 - \frac{ \|y_{exp} - y_{sim}\|}{\|y_{exp} - \overline{y_{exp}}\|}.
\end{aligned}
\end{equation}

\noindent
The NRMSE for each data set ranges from $-\infty$ (poorest fit) to $1.0$ (best fit). For example, Fig.\ \ref{fig:Dim03} shows the $19$ individual radius vs.\ time data sets for 0.3\% gel in dimensional form. The black trace is the simulation obtained using the Neo-Hookean model initialized with a $G$, $\mu$, and $R_0$ that best fit a representative data set (black squares). For this representative case, $G = 8.5$ kPa, $\mu = 0.088$ Pa$\cdot$s, and $R_0 = 1.4$ $\mu$m, and the fit is $0.97$. 

\subsubsection{Data Assimilation Approach}

To account for additional modeling and experimental uncertainties, \citet{Spratt_2020} modified the IMR method of \citet{estrada2018high}, using ensemble-based data assimilation to estimate viscoelastic material properties with the same laser-induced cavitation measurements. This approach is straightforward to apply to the present ultrasound measurements, with  minor modifications. In particular, to verify the above results and examine uncertainty in parameter estimates, we apply the hybrid ensemble-based 4D-Var method (En4D-Var)\citep{Spratt_2020}, which is well-suited to experimental data with smaller numbers of measurements per data set.

En4D-Var is based on the iterative ensemble Kalman smoother \citep{Bocquet_2013,Sakov_2012}, in which the state dynamics are represented through an ensemble of $q$ state vectors, the mean of which represents the estimate at any given time. The state vector $\bm{x}$ is comprised of all the dependent variables, to which the parameters to estimate are appended.
Here, the state vector is
\begin{equation}
    \bm{x} = \{ R, \dot{R}, p_b, 
    S, \mathbf{T}, \mathbf{C}, 
    G, \mu, R_0, \alpha, t_s \},
\end{equation}
where $R$ is the bubble wall radius, $\dot{R}$ the velocity, $p_b$ the bubble pressure, $S$ the stress integral, $\mathbf{T}$ and $\mathbf{C}$ the discretized temperature and vapor concentration fields inside the bubble, $G$ the shear modulus, $\mu$ the viscosity, $R_0$ the stress-free radius, $\alpha$ the stiffening parameter in the QLKV model, and $t_s$ a time-shift parameter used to initialize the bubble growth. The different models (Neo-Hookean and both QLKV models) from section \ref{sec:model} are re-written as nonlinear operators, $F$, which step each state vector in the ensemble forward in time such that $F(\bm{x}_k)=\bm{x}_{k+1}$. A linear operator $H$ is defined, which maps the state vector to measurement space. Here,  the state vector is mapped to its first element $H(\bm{x}_k) = R_k$ as we use radius measurements. Using a Gauss-Newton algorithm, the En4D-Var then minimizes the following cost function:
\begin{equation}
    J(\bm{x}) = \frac{1}{2} \sum_k \beta_k 
    \lVert \bm{y}_{k}-H\circ F_{k \leftarrow 0}(\bm{x})\rVert^2_{\bm{R}} + 
    \frac{1}{2} \lVert \bm{x}-\hat{\bm{x}}_k\rVert^2_{\bm{\mathcal{C}}_0},
\end{equation}
where $\beta_i$ are weights attributed to given time steps, $\bm{y}_k$ is the radius data at time $k$, $\bm{R}$ is the estimated measurement noise covariance matrix, and $\bm{\mathcal{C}}_0$ is the initial ensemble covariance. The first term minimizes the difference with experimental data across the entire time domain, weighed by estimated measurement error, while the second term minimizes difference with the ensemble average, weighed by the ensemble covariance. More details about the algorithm can be found in \citet{Spratt_2020}.

This method is implemented with both gels, using the three material models described in section \ref{sec:inference}. An ensemble size of $q=48$ was used, and the initial ensemble is sampled from a Gaussian distribution centered around an initial state vector. The dependent variables in this initial state vector are taken from the IMR code initialization, and values near IMR results are used as initial guesses for the parameters to estimate.

\section{Results}

\subsection{IMR Results}

\begin{figure*}
\begin{center}
\begin{minipage}{0.90\textwidth}
  \includegraphics[width=\textwidth]{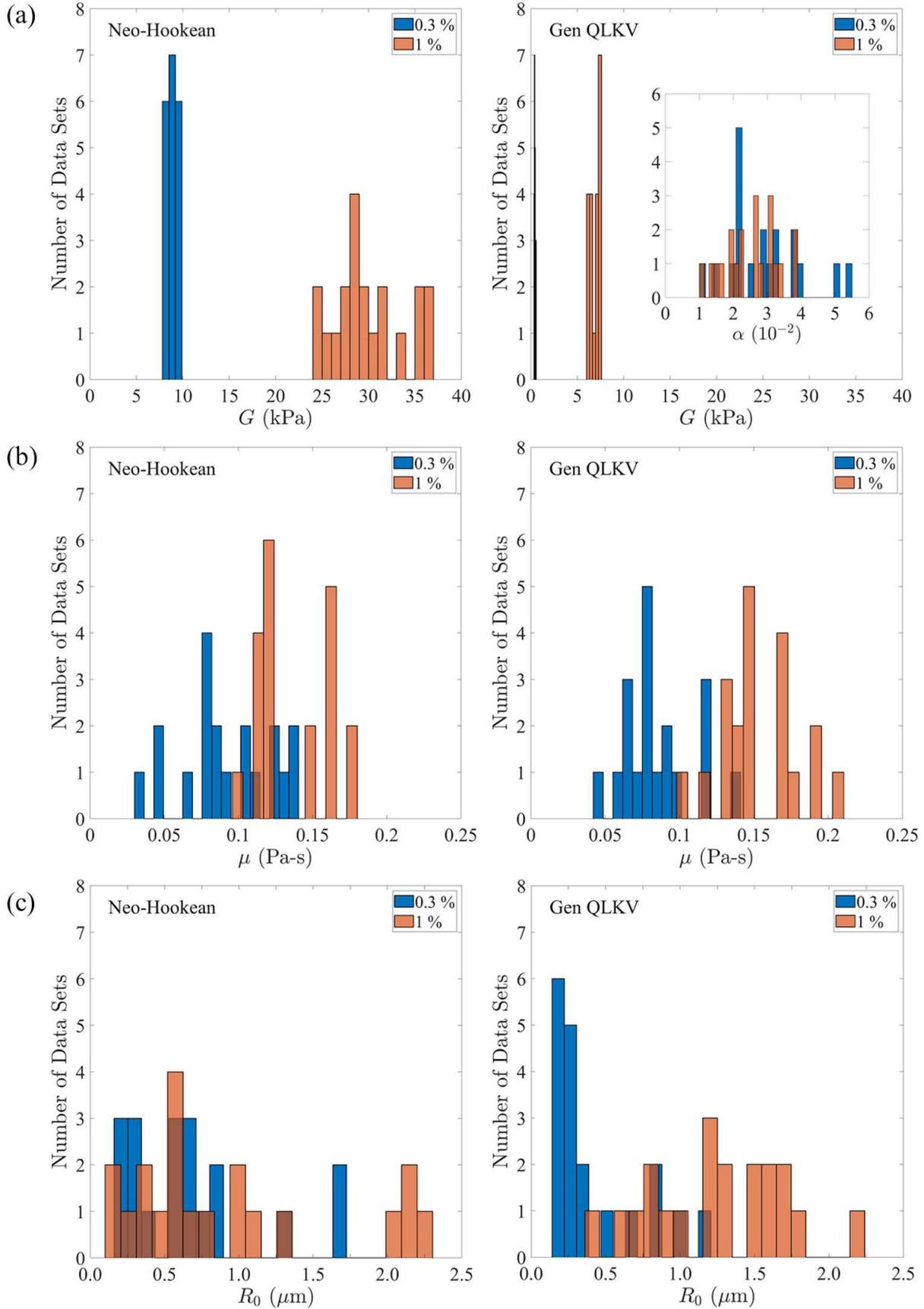}
  \caption{Distributions of (a) shear modulus with stiffening parameter as inset, (b) viscosity, and (c) stress-free radius for $0.3\%$ and $1\%$ agarose gels obtained with Neo-Hookean ($\alpha = 0$) and general QLKV models.}
  \label{fig:Dist}
\end{minipage}
\end{center}
\end{figure*}

For a given experimental data set $(R_i, t_i)$, an ensemble of simulations is run to obtain the time history of the radius, $R(t)$. In each ensemble, the shear modulus, $G$, stiffening parameter,  $\alpha$, viscosity, $\mu$, and stress-free radius, $R_0$, are varied assuming one of three viscoelastic models: Neo-Hookean, quasi-static QLKV, and general QLKV. The parameters in the simulation producing the smallest normalized root-mean squared error constitute the best fit. The mean and standard deviations of IMR results for each parameter weighted by normalized rms error are summarized for each model in Table \ref{table:Props03} for the 0.3\% and 1\% gel data. Figure \ref{fig:Dist} shows the distributions of each parameter obtained with the Neo-Hookean and general QLKV models using IMR. 

The shear modulus distributions differ notably for the Neo-Hookean and QLKV models. In the 0.3\% gel, a weighted mean $G$ of $9.0$ kPa is inferred with the Neo-Hookean model, which is significantly larger than the mean quasi-static measurement of $0.38$ kPa and the weighted mean of $0.42$ kPa inferred with the general QLKV model.  The $G$ distribution for the general QLKV model is also more narrow than that obtained with the Neo-Hookean or quasi-static QLKV models and falls within the 95\% confidence interval of the mean quasi-static measurement. All distributions are broader for the 1\% concentration gel, but the trends are otherwise similar to the 0.3\% gel results. Namely, a large mean $G$ of $30$ kPa is obtained with the Neo-Hookean model. The significantly smaller mean quasi-static measurement of $7.2$ kPa is comparable to the mean of $6.9$ kPa inferred with the general QLKV model. Again, the mean $G$ obtained with the general QLKV model falls within the 95\% confidence interval of the quasi-static measurement. Shear modulus distributions inferred using the Neo-Hookean and QLKV models for each gel concentration are shown in Fig.\ \ref{fig:Dist} (b), which illustrates the smaller  $G$ values and minimal variance achieved with the general QLKV model. The quasi-static and general QLKV models both include an additional elastic constant: the stiffening parameter, $\alpha$. In the 0.3\% gel, a weighted mean $\alpha$ of $0.026$ is obtained with the quasi-static QLKV model. A larger mean $\alpha$ is obtained with the general QLKV model, but the distributions inferred with either model have similar variance. In the 1.0\% gel, the mean values obtained with both QLKV models are $0.025$, but use of the general QLKV model results in a slightly broader $\alpha$ distribution. The $\alpha$ distributions obtained with the general QLKV model are shown in the inset in Fig. \ref{fig:Dist} (b). This distribution highlights the minimal distinction between $\alpha$ distributions inferred from the $0.3\%$ and $1\%$ gel data.

The viscosity distributions are similar for each material model. In the 0.3\% gel, a weighted mean $\mu$ of $0.092$ Pa$\cdot$s is inferred with the Neo-Hookean model while the QLKV models both result in a weighted mean viscosity of $0.086$ Pa$\cdot$s. The distributions also demonstrate similar variance. Inferred viscosity distributions for the 1\% gel have larger means but are again similar for each material model. A weighted mean $\mu$ of $0.14$ Pa$\cdot$s is obtained with the Neo-Hookean model, and both QLKV models produce the same, slightly larger mean $\mu$ of $0.15$ Pa$\cdot$s. Figure \ref{fig:Dist} (c) shows the $\mu$ distributions obtained with the Neo-Hookean and general QLKV models for each gel concentration. Significant overlap between $\mu$ distributions inferred with the Neo-Hookean and general QLKV models is evident. 

The stress-free radius distributions inferred with each material model demonstrate significant overlap.  For the 0.3\% gel data, optimized $R_0$ values obtained with the Neo-Hookean model are centered about a weighted mean of $0.68$ $\mu$m with a standard deviation of $0.45$ $\mu$m. The distribution of $R_0$ values obtained with the quasi-static QLKV model is more narrow but has a comparable weighted mean of $0.43$ $\mu$m. Use of the more general QLKV model with flexible $G$ results in the same mean $R_0$ as in the quasi-static case but with further narrowing of the $R_0$ distribution. All models produce larger $R_0$ values for the 1\% gel. The mean $R_0$ inferred with the Neo-Hookean model is $0.93$ $\mu$m with a standard deviation of $0.66$ $\mu$m. Both the quasi-static and general QLKV models give rise to a larger mean $R_0$ of $1.3$  $\mu$m, but the general QLKV model again gives a more narrow distribution. Stress-free radius histograms obtained with the Neo-Hookean and QLKV models for each gel concentration are shown in Fig. \ref{fig:Dist} (a). Relative to the Neo-Hookean model, use of the general QLKV model results in tighter $R_0$ distributions as well as greater distinction between $R_0$ distributions inferred from the $0.3\%$ and $1\%$ gel data.

\begin{figure*}
\begin{minipage}{\textwidth}
\begin{center}
  \includegraphics[width=\textwidth]{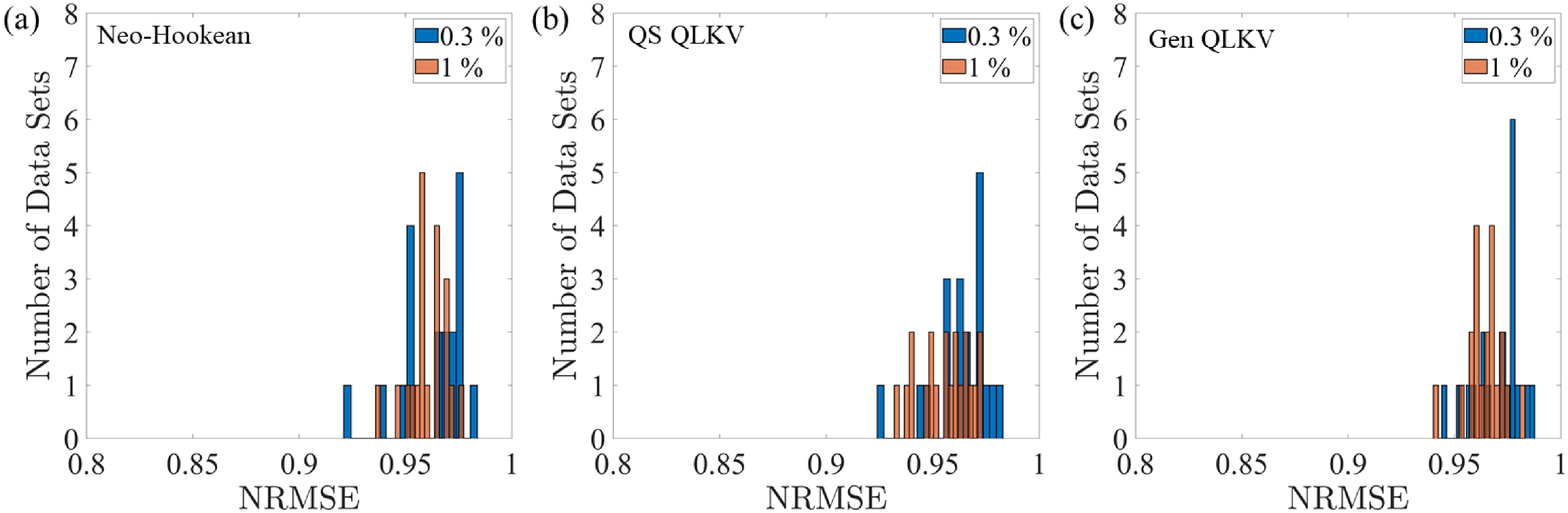}
  \caption{Distributions of normalized rms error (NRMSE) obtained with the (a) Neo-Hookean, (b) quasi-static QLKV, and (c) general QLKV models.}
  \label{fig:errorDist}
  \end{center}
\end{minipage}
\end{figure*}

Normalized rms error (NRMSE) distributions for IMR-optimized parameters obtained using the Neo-Hookean, quasi-static QLKV, and general QLKV models are shown in Figure \ref{fig:errorDist}. NRMSEs $\textgreater 0.9$ are achieved for all data sets, regardless of material model. Both QLKV models achieve NRMSE distributions with comparable means but smaller variance than the Neo-Hookean model. For the $0.3\%$ gel data sets with IMR,  Neo-Hookean model NRMSE ranges from $0.92$ to $0.98$ with a mean of $0.96$, quasi-static QLKV model NRMSE ranges from $0.93$ to $0.98$ with a mean of $0.96$, and general QLKV model NRMSE ranges from $0.94$ to $0.99$ with a mean of $0.97$. Smaller NRMSEs and less distinction between models is seen in the 1\% gel data sets: The Neo-Hookean model and both QLKV models achieve a mean NRMSE of $0.96$. 

\begin{table}[t!]
\small
  \caption{\ Weighted mean and standard deviation of inferred properties for 0.3\% and 1\% agarose specimens obtained using Neo-Hookean (NH), quasi-static QLKV (QS QLKV), and general QLKV (Gen QLKV) models (mean $\pm$ standard deviation). Note that $G$ for the QS QLKV model is a measured value reported as mean $\pm$ 95\% confidence interval. }
  \label{table:Props03}
  \begin{tabular*}{0.49\textwidth}{@{\extracolsep{\fill}}lllll}
    \hline
    Model & $G$ (kPa) & $\alpha$ ($10^{-2}$) & $\mu$ (Pa$\cdot$s) & $R_0$ ($\mu$m)\\
    \hline\hline
    \ 0.3\% gel \\
    \hline
    NH  & 9.0 $\pm$ 0.62 & 0 & 0.092 $\pm$ 0.031 & 0.68 $\pm$ 0.45  \\
    QS QLKV & 0.38 $\pm$ 0.16 &  2.6 $\pm$ 1.2 & 0.086 $\pm$ 0.023 & 0.43 $\pm$ 0.38\\
    Gen QLKV  & 0.42 $\pm$ 0.062 &  2.9 $\pm$ 1.1 & 0.086 $\pm$ 0.024 & 0.43 $\pm$ 0.31\\
    \hline\hline
    \ 1\% gel \\
    \hline
    NH  & 30 $\pm$ 3.8 & 0 & 0.14 $\pm$ 0.025 & 0.93 $\pm$ 0.66\\
    QS QLKV  & 7.2 $\pm$ 0.33 & 2.5 $\pm$ 0.62 & 0.15 $\pm$ 0.022 & 1.3 $\pm$ 0.49 \\
    Gen QLKV  & 6.9 $\pm$ 0.49 & 2.5 $\pm$ 0.77 & 0.15 $\pm$ 0.026 &  1.3 $\pm$ 0.44  \\
    \hline
  \end{tabular*}
\end{table}

\subsection{En4D-Var Results}

We now analyze the simulations results by running them through En4D-Var. Initial guesses as to each material property are determined based on the estimates in Table \ref{table:Props03}. For example, in the 0.3\% gel case, the initial guesses with all models are $R_0 = 0.5~\mu$m and $\mu=0.1$~Pa$\cdot$s. For the Neo-Hookean model, the shear modulus initial guess is $G=10$ kPa. For both QLKV models, the initial guess is $\alpha = 0.03$, and while the shear modulus is fixed at $G=0.38$ kPa in the quasi-static case, the initial guess is $G=0.5$~kPa in the general QLKV case.

\begin{table}[t]
\small
  \caption{\ Weighted mean and standard deviation of inferred properties using En4D-Var for 0.3\% and 1\% agarose specimens. }
  \label{table:en4d}
  \begin{tabular*}{0.48\textwidth}{@{\extracolsep{\fill}}lllll}
    \hline
    Model & $G$ (kPa) & $\alpha$ ($10^{-2}$) & $\mu$ (Pa$\cdot$s) & $R_0$ ($\mu$m)\\
    \hline\hline
    \ 0.3\% gel \\
    \hline
    NH  & 9.66 $\pm$ 0.55  & 0 & 0.086 $\pm$ 0.028 & 0.51 $\pm$ 0.07\\
    QS QLKV & 0.38 $\pm$ 0.16 &  3.0 $\pm$ 0.1 & 0.097 $\pm$ 0.026 & 0.50 $\pm$ 0.03 \\
    Gen QLKV   & 0.50 $\pm$ 0.05 &  3.0 $\pm$ 0.3 & 0.094 $\pm$ 0.026 & 0.51 $\pm$ 0.06\\
    \hline\hline
    \ 1\% gel \\
    \hline
    NH & 36 $\pm$ 3.57 & 0 & 0.14 $\pm$ 0.031 & 1.03 $\pm$ 0.04 \\
    QS QLKV & 7.2 $\pm$ 0.33 & 2.4 $\pm$ 0.15 & 0.16 $\pm$ 0.038 & 1.29 $\pm$ 0.06 \\
    Gen QLKV  & 7.7 $\pm$ 0.89 & 2.5 $\pm$ 0.12 & 0.16 $\pm$ 0.036 &  1.29 $\pm$ 0.05 \\
    \hline
  \end{tabular*}
\end{table}

Table 2 summarizes the results with En4D-Var for both the 0.3\% and 1\% agarose specimens. These results are comparable with those in table \ref{table:Props03}. A key difference, however, appears in the comparatively much smaller stress-free radius and stiffening parameter standard deviations. In fact, the estimates for these parameters are very close to the initial guess in all cases. This indicates that given these initial guesses the En4D-Var is unable improve on these results and converged quickly to the original value. This behavior is further discussed in section \ref{sec:uncertainty}

The radius-normalized RMS errors obtained with the En4D-Var are also similar those obtained with IMR. For example, in the 0.3\% gel case, the normalized RMS errors ranges from 0.91 to 0.98 with a mean of 0.96 with the Neo-Hookean model. For the quasi-static QLKV model, they range from 0.93 to 0.98 with a mean of 0.95. Finally for the general QLKV model, the range is 0.94 to 0.98 with a mean of 0.96.

\section{Discussion}

The present study is the first to use single-bubble acoustic cavitation data to measure the nonlinear viscoelastic properties of soft materials. Like the laser-based IMR method \cite{estrada2018high}, acoustic cavitation rheometry can be used to infer the local mechanical properties of complex soft materials subjected to high strain-rate ($>10^3$ s$^{-1}$) loading. Single-bubble radius vs.\ time measurements are coupled to material properties using bubble dynamics modeling. Property distributions are thus influenced by both model-based and parametric uncertainties. Perhaps the greatest model-based uncertainty lies in the form of constitutive model assumed for the soft matter specimens. We adopt finite-deformation Kelvin-Voigt models in this work given their quantitative success in physics-based modeling of cavitation in soft matter with a limited number of parameters \cite{mancia2019modeling,mancia2017predicting,vlaisavljevich2016visualizing,estrada2018high}. The NRMSE fit distributions in Fig. \ref{fig:errorDist} demonstrate improving data fit and greater fit precision as models increase in complexity from Neo-Hookean to general QLKV. Additional model-based uncertainty remains in the acoustic forcing waveform chosen for our numerical simulations. The precise, time-dependent acoustic forcing experienced by bubbles in the focal region cannot be directly measured due to damaging cavitation at the hydrophone tip. Our analytic approximation is successful in high-amplitude ultrasound contexts \cite{mancia2020measurements,mancia2020single} and is a reasonable choice of pressure waveform in these water-based gels observed to have the same cavitation threshold \cite{wilson2019comparative}. A prior validation study in water supports the validity of this analytic approximation for high-amplitude acoustic forcing but concedes that dedicated experiments are necessary to determine the exact pressure experienced the bubble \cite{mancia2020single}. Parametric uncertainties in addition to the stress-free radius, elastic constants, and viscosity values measured by this method include material constants such as surface tension and thermodynamic parameters, which are assumed equivalent to their values for water \cite{mancia2020single}. This assumption is appropriate for these water-based gels but variation of these parameters should be considered if this method is used to characterize more complex materials. Each of the material properties inferred for the agarose gel specimens in this study are now discussed in greater detail.

\subsection{Elastic Parameters}
Shear modulus is inferred using an assumed material model, which we take to be a Neo-Hookean, quasi-static QLKV, or general QLKV model.  \citet{vlaisavljevich2015effects} measured the Young's moduli for agarose gels with a parallel plate rheometer; these moduli were subsequently used as physical parameters for simulations of cavitation in agarose gels \cite{wilson2019comparative,vlaisavljevich2015effects}. Agarose is considered nearly incompressible at high strain rate \cite{normand2000new}, so these values can be readily converted to mean shear moduli of $0.38$ kPa for $0.3\%$ gel and $7.2$ kPa for $1\%$ gel. Notably, these measurements were performed on gross specimens under quasi-static conditions.  In general, local shear moduli inferred from inertial cavitation experiments are expected to be larger given the stiffening behavior observed in gels subjected to high strain-rates \cite{wang2016dynamic,kwon2010compressive}. The Neo-Hookean model infers local mean shear moduli that are more than 4 times larger than these quasi-static measurements for each gel specimen. Similarly, application of IMR to polyacrylamide specimens using laser-induced cavitation data and a Neo-Hookean model yielded shear moduli at least two times larger than quasi-static measurements \cite{estrada2018high}.   

The quasi-static and general QLKV models include gel shear modulus and introduce another elastic parameter: the stiffening constant, $\alpha$. Our results show minimal distinction between the parameter distributions obtained with either QLKV model, which shows that even a QLKV model with variable shear modulus effectively infers the measured quasi-static shear modulus for each gel specimen. Values for $\alpha$ in the QLKV models are fairly similar for gel specimens of either concentration, but trend slightly larger for the $0.3\%$ gel. A trend of smaller $\alpha$ for stiffer samples was observed previously in a cavitation rheometry study of soft and stiff polyacrylamide \cite{yang2020extracting}. This polyacrylamide study also measured $\alpha$ values of $0.5-1.0$, which are significantly larger than those in agarose. This contrast likely reflects the distinct microstructures of agarose and polyacrylamide. The significantly larger shear moduli predicted by the Neo-Hookean model relative to the QLKV model are a direct consequence of neglecting higher-order stiffening effects captured with the $\alpha$-dependent terms of Eq. \ref{eq:SS}. Inclusion of these effects is also responsible for the greater fit accuracy and precision achieved with the QLKV models. 

\subsection{Viscosity}
Viscosity of the gel specimens is also inferred and shown to be largely independent of the assumed viscoelastic model. \citet{wilson2019comparative} proposed using an agarose viscosity of $0.115$ Pa$\cdot$s for $0.3\%$, $1\%$, $2.5\%$, and $5\%$ gel specimens \cite{wilson2019comparative}. Although they admit this value is uncertain, it was found to result in a relationship between initial radii and gel concentration that followed the same approximate scaling as agarose pore size and gel concentration.  As noted by previous authors \cite{movahed2016cavitation}, there are currently no measured values of agarose viscosity relevant to cavitation conditions. Past studies of ultrasound-induced cavitation in agarose have assumed water viscosity ($0.001$ Pa$\cdot$s) \cite{vlaisavljevich2015effects} or considered a range of viscosities ($0.001$ - $10$ Pa$\cdot$s) \cite{movahed2016cavitation} to account for this otherwise unknown quantity. Meanwhile, shear wave elastography techniques have been used to infer gel viscosity under more typical conditions. For example, a prior study assuming a Kelvin-Voigt type material model measured a viscosity of $0.22$ Pa$\cdot$s for an agar-gelatin phantom at $400$ kHz \cite{catheline2004measurement}. Still assuming a Kelvin-Voigt model, other authors measured a viscosity of $1$ Pa$\cdot$s under ballistic loading of a $10$ wt\% ballistic gelatin block \cite{liu2014viscoelastic}. Larger viscosity values of $5$ to $900$ Pa$\cdot$s have also been measured for agar under frequencies ranging from $20$ to $200$ Hz \cite{nayar2012elastic}. In part due to this wide variation in measurements, \citet{movahed2016cavitation} conclude that a single viscosity parameter cannot fully describe dissipative behavior of the gels but that the effective viscosity of agarose should be assumed to be larger than that of water. Our findings support the use of viscosity values on the order of $0.1$ Pa$\cdot$s for agarose gels, which is also consistent with the polyacrylamide viscosities ($0.101$ Pa$\cdot$s and $0.118$ Pa$\cdot$s) obtained from laser-induced cavitation data and a Neo-Hookean model \cite{estrada2018high}.

\subsection{Stress-Free Radius}

Our use of acoustic rather than laser-induced cavitation experiments requires the consideration of an additional parameter, the stress-free radius, $R_0$, which is equivalent to the initial radius in our simulations. Compared to the conventional laser-based IMR procedure, we fit the entire bubble growth and collapse, not just the latter, thus bypassing the need to assume a given state (temperature and composition fields inside the bubble) at the bubble's maximum size. Instead, an initial (stress-free) radius is provided. Numerous studies have suggested that cavitation bubbles in high-amplitude ultrasound arise from pre-existing nuclei \cite{mancia2020measurements,wilson2019comparative,bader2019whom,vlaisavljevich2016visualizing,vlaisavljevich2016effects,vlaisavljevich2015effects,vlaisavljevich2014histotripsy,maxwell2013probability}. Previous work has used high-amplitude ultrasound experiments to infer the nucleus size distribution at the acoustic cavitation threshold in water \cite{mancia2020measurements}. In the setting of agarose gels and other soft matter, the physical significance of the initial radius parameter is less clear, but acoustic cavitation in soft materials likely originates from pre-existing defects which act as stress risers and are analogous to cavitation nuclei in liquids \cite{guan2013cavitation,wilson2019comparative}. \citet{wilson2019comparative} first hypothesized that stress-free radii could be related to agarose porosity. They found a correlation between measured agarose pore size \cite{narayanan2006determination,pernodet1997pore} and initial radii they estimated for representative agarose gel specimens of varying concentration. Using only maximum radii of three data sets per gel specimen, they inferred nuclei sizes of approximately $1.4$ $\mu$m in $0.3\%$ gel and $0.25$ $\mu$m in $1\%$ gel. In contrast, our present study finds that stress-free radii are larger in the higher concentration gel, with inferred sizes of approximately $0.43$ $\mu$m in $0.3\%$ gel and $1.3$ $\mu$m in $1\%$ gel. These values could be a reflection of where cavitation occurs in both gel specimens.  For example, the stress-free radius distribution for $0.3\%$ gel is weighted towards smaller values, which suggests bubbles might be arising from nanoscale nuclei \cite{mancia2020measurements} contained in water-filled pores. In this case, bubbles grow from nanometer-sized nuclei but are affected by elasticity only after reaching micron-scale sizes, which may be on the order of the pore size. In addition, fewer cavitation events occur in the $1\%$ gel, implying that experiments could be preferentially nucleating the largest pores in this stiffer gel specimen. Finally, the assumption that agarose pore size is related to the size of cavitation nucleus is reasonable but unproven. An alternative explanation is that gels immediately fracture under the high stresses and strains developed at the onset of cavitation \cite{mancia2019modeling}. This process could then lead to `effective' initial or stress-free radii sizes which have no clear correlation with pore size. 

\subsection{Uncertainties} \label{sec:uncertainty}

The En4D-Var results can further inform the uncertainties associated with the material property estimates obtained, e.g., due to variations in samples.

\begin{figure}[!t]
\centering
  \includegraphics[width=0.48\textwidth]{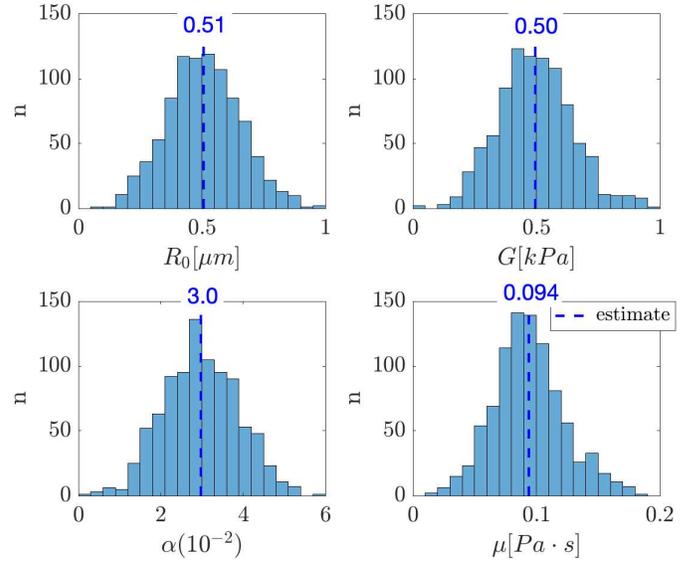}
  \caption{Histogram of combined final stress-free radius, shear modulus, stiffening parameter and viscosity ensembles for $0.3\%$ gels with the En4D-var using the QLKV model. }
  \label{fig:En4D_hist_Fung}
\end{figure}

\noindent
Figure \ref{fig:En4D_hist_Fung} summarizes the En4D results with the QLKV model for each material property. These histograms combine all final ensemble members across the 19 data sets (with an ensemble size of 48, the total number of ensemble members is thus $19 \times 48 = 912$). An approximately Gaussian distribution is obtained for all four quantities, the mean of which are our estimates for each quantity, thus confirming that the En4D-Var estimates are uniform across all data sets.

\begin{figure}[!t]
\centering
  \includegraphics[width=0.48\textwidth]{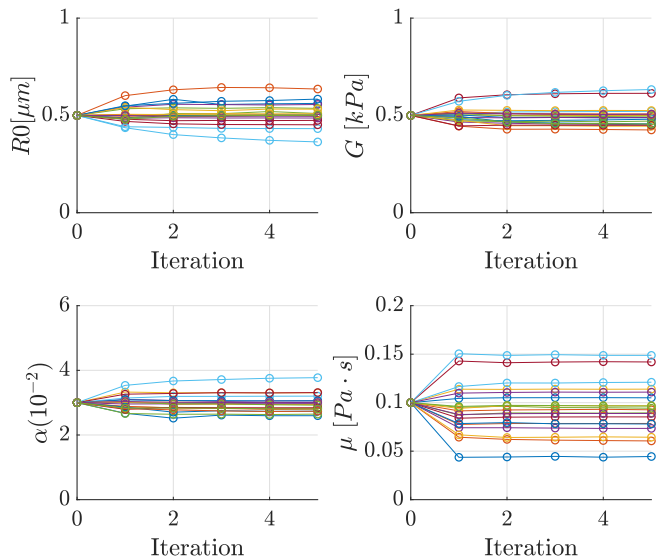}
  \caption{Iterative estimation of the stress-free radius, shear modulus, stiffening parameter and viscosity for $0.3\%$ gels with the En4D-var using the QLKV model. }
  \label{fig:En4D_Fung}
\end{figure}

Figure \ref{fig:En4D_Fung} shows the iterative estimation of each parameter for all 19 data sets, again with the QLKV model. It appears that the viscosity estimates are relatively scattered, which points to a relatively high sensitivity to changes in viscosity in our estimator. By contrast, the spread in stress-free radius, shear modulus and stiffening parameter is quite narrow around the initial guess, indicating that in this regime and given this guess the estimator could not improve the fit significantly. Overall, this uncertainty assessment demonstrates that the acoustic cavitation extension of IMR applied with data assimilation methods provides robust parameter estimates comparable to those obtained with the traditional IMR optimization approach.

\subsection{Acoustic vs. laser-induced cavitation data}
This work demonstrates that the IMR method can be applied using radius vs.\ time data from ultrasound-nucleated cavitation bubbles. Novel experimental techniques \cite{wilson2019comparative} and recent validation of models for bubble dynamics under high-amplitude ultrasound forcing permit this extension of the IMR method. Notably, the latter validation studies \cite{mancia2020single} enable characterization of the high-amplitude acoustic waveform, a previously significant source of model-based uncertainty. Use of acoustic cavitation measurements removes uncertainties associated with laser-material interactions and optical breakdown.  Acoustic cavitation also has direct relevance to clinical ultrasound applications and could more closely approximate cavitation phenomena in blast injuries. Laser cavitation measurements are still advantageous in avoiding uncertainties associated with the acoustic forcing waveform and with the likely stochastic distribution of pre-existing cavitation nuclei and associated stress-free radii. At the present time, laser experiments are more robust and repeatable than ultrasound-based experiments, in that a single bubble is nucleated, grows, and collapses, by contrast to acoustically generated bubbles sometimes breaking up into smaller bubbles before any rebounds are observed \cite{duryea2015removal}. However, it is conceivable that the advantages and disadvantages of each method could ultimately prove complementary. For instance, a combined approach could involve use of laser-induced cavitation data to determine cavitation-relevant material parameters, followed by use of acoustic cavitation data to determine local waveform characteristics. A similar approach has been used to determine mechanical properties of gels at lower rates using continuous pressure waveforms \cite{oguri2018cavitation,shirota2015estimation}.

\section{Conclusions}

Cavitation-based rheometry techniques provide a minimally invasive means of characterizing soft, viscoelastic materials such as gelatin and biological tissue.  This work presents a novel cavitation rheometry technique using radius vs. time data obtained from acoustic cavitation experiments. Based on focused ultrasound radius vs. time data and using a numerical model for single bubble dynamics in a finite deformation Kelvin-Voigt medium with either a Neo-Hookean or higher order Quadratic Law elastic term, we infer properties including stress-free radius, elastic parameters, and viscosity of $0.3\%$ and $1\%$ agarose gel specimens first studied by \citet{wilson2019comparative}. Our findings illustrate the utility of single-bubble acoustic cavitation for measurement of viscoelastic properties. Use of acoustic cavitation data is advantageous in avoiding the complications of optical breakdown and potential material property alterations in laser-induced cavitation. Acoustic cavitation rheometry is ideally suited for inference of tissue properties in the setting of high-amplitude ultrasound treatments. Furthermore, the presented acoustic cavitation extension of IMR is shown to be robust, obtaining comparable parameter values when used in conjunction with novel data assimilation methods. 

\section*{Conflicts of Interest}
There are no conflicts to declare.

\section*{Acknowledgements}
This work was supported by ONR Grant No. N00014-18-1-2625  (under  Dr.\ Timothy  Bentley).



\balance

\bibliography{rsc} 
\bibliographystyle{rsc} 

\end{document}